\documentclass[aps,notitlepage,a4paper,superscriptaddress,11pt]{article}

\usepackage[utf8]{inputenc}
\usepackage[dvips]{color}
\usepackage[toc]{appendix}
\usepackage[hang,flushmargin]{footmisc} 
\usepackage{titlesec,titletoc,ntheorem-hyper,authblk,abstract,latexsym,microtype,booktabs,amsmath,cleveref,fancyhdr,wrapfig,array,amsfonts, amssymb,geometry,appendix,cite}
\usepackage{dsfont}
\usepackage{amsmath, amssymb, graphics, setspace}

\newcommand{\be}{\begin{equation}}
\newcommand{\ee}{\end{equation}}
\newcommand{\bea}{\begin{eqnarray}}
\newcommand{\eea}{\end{eqnarray}}
\newcommand\blfootnote[1]{%
  \begingroup
  \renewcommand\thefootnote{}\footnote{#1}%
  \addtocounter{footnote}{-1}%
  \endgroup
}

\usepackage{caption}
\usepackage{subcaption}
\usepackage{amsmath}
\usepackage{amsfonts}
\usepackage{amssymb}
\usepackage{mathtools}
\numberwithin{equation}{section}
\usepackage{titlesec}
\usepackage{sectsty}
\sectionfont{\centering}
\usepackage{framed}
\allowdisplaybreaks
\usepackage[nottoc]{tocbibind}
\usepackage[free-standing-units]{siunitx}
\usepackage{helvet}
\usepackage{url}
\usepackage{titletoc}
\usepackage{blindtext}
\usepackage[gen]{eurosym}

\title{Exceptional point in a coupled Swanson system}
\author{Bijan Bagchi, Rahul Ghosh, and Sauvik Sen}
\affil{Physics Department, Shiv Nadar University, Gautam Buddha Nagar, \\
Uttar Pradesh 203207, India}

\begin{document}

\maketitle

\begin{abstract}
Abstract: We propose an interacting nonhermitian model described by a two-mode quadratic Hamiltonian along with an interaction term to locate and analyse the presence of an exceptional point in the system. Each mode is guided by a Swanson-like quadratic Hamiltonian and a suitable choice is made for the interaction term. The parity-time symmetric transformation is adopted in the standard way relevant for a coupled system.
\end{abstract}

\blfootnote{E-mails: bbagchi123@gmail.com, rg928@snu.edu.in, ss128@snu.edu.in}

\section{Introduction}

Nonhermitian Hamiltonians undergo non-unitary evolution and generally represent open quantum systems with balanced gain and loss when interacting with the environment \cite{moi, rot1}. The subclass embodying $\mathcal{PT}$-symmetry, introduced by Bender and Boettcher about a couple of decades ago, has proved to be a fertile ground for intense inquiry as exemplified by outpouring of research outputs in this direction \cite{ben1, ben2, chr, baga}. $\mathcal{PT}$-symmetric Hamiltonians may support real or complex
conjugate pairs of energy eigenvalues under certain conditions related to such a symmetry being exact or
spontaneously broken. However, unlike Hermiticity which generally governs a closed and isolated system accompanied by unitary time evolution, and serves as a sufficient condition for the Hamiltonian to possess real eigenvalues, $\mathcal{PT}$-symmetry is neither necessary nor sufficient to ensure reality of the spectra. Exceptional points (EPs) may occur as critical values where symmetry breaking takes place and are regarded as a typical feature of nonhermitian systems \cite{kat, heiss, ply}. These unconventional points refer to solutions of the parameter equation at which the associated
eigenvectors are parallel. Basically, an EP is a branch point singularity where two or more discrete eigenvalues, corresponding to two different quantum states, along with their accompanying eigenfunctions coalesce (for nice introductions to the subject, see \cite{zno, fer1}) pointing to degeneracies in the system. The latter ceases to obey any conservation law because of its open character \cite{kat}. Indeed, while approaching the exceptional point, the phases of the eigenfunctions of the Hamiltonian which is nonhermitian, may not show sufficient robustness, as a result information from outside can get leaked into the system\cite{rot2, yu}. 

Potential appearance in a wide range of physical problems like those in optical and condensed matter models, coupled cavity problems and photonics where high resolution control of optical gain and loss is possible, has sparked interest in systems exhibiting EPs \cite{mir, dow}. In particular, observations of phase transitions and occurrence of EPs as a three-state coalescence in optical trimers \cite{jin}, wherein a linear chain of three coupled cavities are connected in an array to have an inbuilt $\mathcal{PT}$-symmetry with adjustable control parameters \cite{fel}, reflect presence of EPs and crossing of it as coupling parameters are varied \cite{xue}. Direct coupling of cavities has led to research in $\mathcal{PT}$-induced enhanced nonlinear characteristics \cite{ram, du} where the behaviour around the EP provides insights to the sensitive coupling dependence at a specific degenerate point. Apart from studies in optical trimer structure \cite{dua, suc1, li, suc2}, works have also been reported for optical dimer involving two coupled cavities \cite{bay, rak} and coupled optical parametric oscillators \cite{roy}. Realization of EPs in other systems include cooling of qubits and interacting oscillators \cite{riv}.  Although not of direct concern to the spirit of the present work, it may be worthwhile to note that since EPs induce quickest transfer of dissipative energy, they have potential relevance in optimal cooling of qubits \cite{part}. Very recently, the  Hopfield-Bogoliubov matrix has been explored to demonstrate the feasibility of EPs for wide subcases of the controlling discriminant \cite{xie}. The singularities at EPs often reveal abnormal topological structures as observed experimentally \cite{den} especially in connection with the topological structures of a Riemann surface \cite{ryu}.  The phase diagrams often help to classify the behavior of the system into different phases that are dependent on the class of external parameters. It is interesting to point out that recent works \cite{zhong} have focused on the Riemann surfaces for the real and imaginary components of the eigenvalues enabling one to track the $\mathcal{PT}$ phases. Curved lines of EPs are found to separate the different phases. With the variation of the parameter the behavior of the system undergo changes because of the trajectory variation in the parameter space.

Interest in $\mathcal{PT}$-systems has led to studies in many coupled systems as well. Encountering EPs in them is a well recognized feature (see \cite{ben3, kha, ben4} and references therein). However, to the best of our knowledge, no studies have so far been undertaken on the Swanson oscillator \cite{swa} which is an extended harmonic oscillator model but designed to be nonhermitian. It was originally proposed to track down transitions of probability amplitudes that are governed by nonunitary time evolution. Motivated by the previous findings, we undertake in this communication a search for the existence of an EP by coupling two harmonic oscillators in an interacting nonhermitian framework and using Swanson's Hamiltonian \cite{swa} as a basis for each of the two oscillators.  

 The model Hamiltonian of Swanson is expressed in terms of
the usual harmonic oscillator creation and annihilation operators $a^\dagger$ and $a$, satisfying the commutation relation $[a, a^\dagger] = 1$,
namely 

\begin{equation}
H^{(  \xi  ,   \chi   )} =  \left (H^{(  \chi   ,   \xi  )} \right)^\dagger = \omega a^\dagger a +   \xi   a^2 +   \chi    {a^\dagger}^2 + \frac{1}{2}\omega
\end{equation}
where $\omega,   \xi  ,   \chi    \in \Re $,
and $\omega,   \xi  ,   \chi   $, such that $  \xi   \neq   \chi   $ and $\Omega^2 = \omega^2 - 4  \xi     \chi    > 0$. It is well known that $H^{(  \xi  ,   \chi   )}$ can be completely solved using standard operator techniques and seen to possess real, positive and discrete spectrum in line
with the conjecture in \cite{ben1}.
Swanson Hamiltonian is obviously
Hermitian but only if $  \xi   =   \chi   $; however, it is $\mathcal{PT}$-symmetric (or, equivalently, $\mathcal{P}$-pseudo-Hermitian \cite{mos}) for all
values of $  \xi  $ and $  \chi   $. This can be readily checked by applying the transformation
properties of $\mathcal{P} : a \rightarrow -a$ and $\mathcal{T} : a \rightarrow a$.  

The Hamiltonian (1.1 is well studied by many authors (see for instance, \cite{fri}).  Jones \cite{jones} showed, by means of a similarity transformation, that $H^{(  \xi  ,   \chi   )}$ admits of an equivalent Hermitian representation (see also \cite{mus, que}). It was also demonstrated to be pseudo-Hermitian \cite{bag} and subsequently a suitable generalization for it was made \cite{ortiz}. A study of non-$\mathcal{PT}$-symmetry phase transition reveals the presence of exceptional points of infinite order \cite{fern}.

We now proceed to deal with an interacting Swanson Hamiltonian to ascertain its phase transitions as one varies the parameters present in it. This is distinguished by specific inequalities among the parameters. An EP appears at the point of phase transition.


\section{Coupled two-mode oscillators}

In connection with the system of two optical resonators \cite{pen}, an algebraic treatment was developed for $\mathcal{PT}$-symmetric coupled bosonic oscillators to find out the classical and quantum mechanical relevance \cite{fer2}, and stability tests were undertaken with regard to the gain-loss parameter \cite{tsoy}. Here we focus on a single pair of Swanson-like  oscillators and label them by 1 and 2. The Hamiltonian ($H^s$) for the union of 1 and 2 is given by

\begin{equation}
    H^s = \mathcal{H}^{(1)} + \mathcal{H}^{(2)}
\end{equation}
Before we come to the coupling of the two oscillators let us note that for the system we are considering we take the following choice of $\mathcal{H}^{(i)}$ 

\begin{equation} \label{shalpha}
    \mathcal{H}^{(i)} = \omega \left (a_i^{\dagger} a_i +\frac{1}{2} \right ) -  \lambda  \left(a_i^2 - a_i^{\dagger^2}\right ), \quad i = 1, 2
\end{equation}
in which $\lambda$ is a positive real parameter. A comparison with (1.1) shows that we have set the corresponding $\omega_1 = \omega_2 = \omega$. Further, in each of $\mathcal{H}^{(i)}$ the counterparts of the coupling constants $  \xi  $ and $  \chi   $ are taken to equal but with differing signs. The case when these are of same sign is suitable for squeezed harmonic oscillator \cite{zha, lo, kry}. In ($\ref{shalpha}$).  Taking interactions into consideration we will have, in general, a combined system of two coupled oscillators. We provide such a facility in the governing Hamiltonian by allowing for the presence of different elements with imaginary coefficients. 

To examine the $\mathcal{PT}$-symmetry of the system (2.1) let us write $a_1 = \frac{1}{\sqrt{2}}(x + ip)$ and $a_2 = \frac{1}{\sqrt{2}}(y + iq)$ and follow  Bender et al's observation \cite{ben4} that under $\mathcal{P}$  

\begin{equation}
\mathcal{P}: x \rightarrow -y, \quad y \rightarrow -x, \quad p \rightarrow -q, \quad q \rightarrow -p
\end{equation}
while under $\mathcal{T}$ 

\begin{equation}
\mathcal{T}: x \rightarrow x, \quad y \rightarrow y, \quad p \rightarrow -p, \quad q \rightarrow -q, \quad i \rightarrow -i
\end{equation}
These reflect that under $\mathcal{PT}$ 

\begin{equation}
\mathcal{PT}: a_1 \rightarrow -a_2^\dagger, \quad a_2 \rightarrow -a_1^\dagger, \quad a_1^\dagger \rightarrow -a_2, \quad a_2^\dagger \rightarrow -a_1
\end{equation}
Applying (2.5) to (2.2) we see that the term involving $\lambda$ changes sign under $\mathcal{PT}$ resulting in $\mathcal{H}^{(1)}$ not transforming to $\mathcal{H}^{(2)}$ and vice versa. Hence the loss and gain
oscillators are not interchanged.

We now consider the impact of  complex frequencies $\omega \pm i\gamma$, signifying oscillatory behaviour in the system, $\gamma$ being a real parameter, and vary the coupling parameters continuously in the parameter space. Our search for an EP lies in determining the transition points and we inquire specifically into the case when the eigenvalues become degenerate and the eigenstates at such a point coalesce together. We identify such a location as a candidate for the EP by following the standard definition of its existence. 

Let us express the total Hamiltonian $H$
as a superposition of the pieces $\mathcal{H}^{(1)}, \mathcal{H}^{(2)}$ along with an interaction term $\mathcal{H}^{(int)}$ that is, $H = \mathcal{H}^{(1)} + \mathcal{H}^{(2)} + \mathcal{H}^{(int)}$, where

\begin{eqnarray}
&& \mathcal{H}^{(1)} = 2(\omega - i\gamma) \left (a_1^{\dagger} a_1 +\frac{1}{2} \right ) -  \lambda  \left(a_1^2 - a_1^{\dagger^2}\right ),  \\
&&\mathcal{H}^{(2)} = 2(\omega + i\gamma) \left (a_2^{\dagger} a_2 +\frac{1}{2} \right ) -  \lambda \left(a_2^2 - a_2^{\dagger^2}\right ),  \\
  && \mathcal{H}^{(int)}  =  \alpha (a_1^\dagger a_2+ h.c)+  \beta (a_1 a_2^\dagger+h.c)+ 2\delta(i a_1^\dagger a_2^\dagger+h.c)
\end{eqnarray}
is defined in terms of non-zero real parameters $ \alpha ,  \beta , \delta$. Note that the interacting Hamiltonian $\mathcal{H}^{(int)}$ is $\mathcal{PT}$-symmetric under the transformations specified in (2.5) but the total Hamiltonian $H$ is not so due to $\mathcal{H}^{(1)}$ not being swapped to $\mathcal{H}^{(2)}$ and conversely. We now proceed to enquire whether our system defined by H exhibits an EP.

We can arrange $H$ in the form $\left (a_1^\dagger \quad a_1 \quad a_2^\dagger \quad  a_2 \right ) M \left (a_1 \quad a_1^{\dagger} \quad a_2 \quad  a_2^{\dagger} \right )^T$ by sandwiching a matrix M between the column $\left (a_1 \quad a_1^{\dagger} \quad a_2 \quad  a_2^{\dagger} \right )^T$ and its hermitian conjugate, with $M$ given by the $4 \times 4$ structure

\begin{equation}
M =	\begin{bmatrix} 
	\omega - i\gamma &  \lambda  &  \alpha  & i\delta \\
	- \lambda  & \omega - i\gamma & -i\delta &  \beta  \\
	 \alpha  & i\delta & \omega + i\gamma &  \lambda  \\
	-i\delta &  \beta  & - \lambda  & \omega + i\gamma
	\end{bmatrix}
	\quad
\end{equation}
Note that $M \neq M^\dagger$. In the analysis of quantum phase transitions, different forms have been chosen for the matrix $M$ \cite{xue, du, xie}.

%

The matrix M provides the indicial equation $\left |M-\Lambda I \right | = 0$ which we solve for the eigenvalues of $M$. When expanded it reads

\begin{eqnarray}
&& \Lambda^4 +j\Lambda^3+k\Lambda^2+l\Lambda+m=0  
\end{eqnarray}
where the coefficients are 
\begin{gather}
 j =-4\omega       \nonumber\\
 k=2\gamma^2-2 \delta^2- \alpha ^2- \beta ^2+2  \lambda  ^2+6 \omega^2          \nonumber\\
 l=4i\delta( \beta + \alpha ) \lambda +2\omega( \beta ^2+ \alpha ^2)-4\omega( \lambda ^2+\gamma^2-\delta^2)-4\omega^3               \nonumber\\
 m= \lambda ^4+\omega^4+\gamma^4+\delta^4+ \alpha ^2 \beta ^2+2 \alpha  \beta  \lambda ^2-4i( \alpha + \beta )\delta\omega \lambda +2 \lambda ^2\omega^2-\omega^2( \alpha ^2+ \beta ^2) \\ -\gamma^2(2\delta^2+ \beta ^2+ \alpha ^2+2 \lambda ^2-2\omega^2)-2\delta^2( \alpha  \beta - \lambda ^2+\omega^2)         \nonumber
\end{gather}
In this work we are interested in exploring the exceptional points arising as a result of a coalescence of different eigenvalues of $M$. These are qualitatively different from the degeneracy character of the Hamiltonian $H$. A thorough theoretical study of (2.10) subject to (2.11) will require an elaborate computational treatment. This and the connection
between the EPs of $M$ and the degeneracy emerging in $H$ will be taken up in a future communication. To keep things tractable and facilitate simple mathematical analysis, we focus on the special case of $ \beta =- \alpha $ when the set of equations (2.11) reduces a great deal and we are left with the parameters $\delta$ and $\alpha$ apart from $ \gamma $ in the matrix M to track down the presence of a plausible EP in the system. It needs to be mentioned that for our adopted choice, the interacting piece of the Hamiltonian will contain only the coupling terms $a_1^\dagger a_2 + \mbox{h.c.}$ but the parameter $\alpha$ continues to be present in $M$ whose variation is exploited to locate the presence of an exceptional point. We proceed to explore this feature in the following section.


In the present scenario it needs to be emphasized that other choices of parameters can well be made enabling to locate more number of EPs. But that is of academic interest only. Our main plan is to demonstrate in a simple scenario how one can run into an EP for a system of coupled oscillators as guided by a specific model of the interacting Swanson type.

\section{Results and discussions}

The eigenvalues $E_{1,2,3,4}$ are the roots of the polynomial equation (2.10) and correspond to the eigenvalues of $M$. These are straightforwardly obtained in the form
\begin{eqnarray}
&& E_1 = \omega - \sqrt{-\gamma^2 + \delta^2 +  \alpha ^2 - \lambda ^2 + 2 \sqrt{\gamma^2  \lambda ^2 - \delta^2  \lambda ^2}} \\
&& E_2 = \omega + \sqrt{-\gamma^2 + \delta^2 +  \alpha ^2 - \lambda ^2 +2 \sqrt{\gamma^2  \lambda ^2 - \delta^2  \lambda ^2}} \\
&& E_3 = \omega - \sqrt{-\gamma^2 + \delta^2 +  \alpha ^2 - \lambda ^2 -2 \sqrt{\gamma^2  \lambda ^2 - \delta^2  \lambda ^2}} \\
&& E_4 = \omega + \sqrt{-\gamma^2 + \delta^2 +  \alpha ^2 - \lambda ^2 -2 \sqrt{\gamma^2  \lambda ^2 - \delta^2  \lambda ^2}} 
\end{eqnarray}
expressed in terms of the given parameters. Observe that all the parameters appear quadratically in $(3.1)-(3.4)$. In principle there should be a relationship between the eigenvalues of $M$ and those of $H$. Concerning the eigenvalues of the latter, let us note that for the nonhermitian quadratic Hamiltonian (2.6) and (2.7) possessing respectively the complex coefficients $\omega - i\gamma$ and $\omega + i\gamma$, the eigenvalues are extremely difficult to construct in a closed form. For the condition $\Omega^2 \geq 0)$ when there are real and positive eigenvalues, Swanson \cite{swa} outlined a procedure to determine exactly the eigenvalues for the case $\gamma = 0$ by means of a generalized Bogoliubov transformation and thereby obtaining a connection with the standard harmonic oscillator. In the process one had to deal with Hilbert space with indefinite metric and non-normalizable eigenstates. The presence of the interaction Hamiltonian $\mathcal{H}^{(int)}$ in the present scenario makes the computation of the eigenvalues of the entire Hamiltonian $H$ even more difficult that takes us beyond the spirit of this work.

To access an EP, we address the expressions of $E_1$ and $E_2$ first and note that both of them converge to the unique single value $\omega$ provided either of the following conditions 

\begin{equation}
    \delta_{(-)} = \pm \sqrt{\gamma^2 - \alpha ^2 - 2 \alpha  \lambda  -  \lambda ^2}
\end{equation}
where $\delta_{(-)}$ is defined with respect to the negative-sign of $2 \alpha  \lambda  $ inside the square-root, are met. However, because of the opposite signs in front of the inner square root of (3.3) and (3.4) as compared to those in (3.1) and (3.2), $E_3$ and $E_4$ remain nondegenerate. To sum up, where the EP occurs, $E_{1,2,3,4}$ are given by

\begin{equation}
E_{1,2,3,4}= \left ({\omega, \omega, \omega - 2\sqrt{- \alpha   \lambda  -  \lambda ^2} ,\hspace{1mm}\omega + 2\sqrt{- \alpha   \lambda  -  \lambda ^2}} \right )
\end{equation}
Interestingly, the sensitive dependence on the parameters $ \alpha $ and $ \lambda $ is revealed when we look at $ \alpha  = -  \lambda $. While for  $ \alpha  < - \lambda $, $E_3$ and $E_4$ are real quantities, for $ \alpha  >  \lambda  $ they transform to a pair of conjugate complex numbers. Exactly at $ \alpha  =  \lambda $ all the four eigenvalues $E_{1,2,3,4}$ merge and coalesce to a single degenerate real value $\omega$. It can be looked as a point of localization where the system undergoes a spontaneous transition from the existing nondegeneracy of $E_3$ and $E_4$ to a collapse into a state of degeneracy of all eigenvalues and subsequently emerging into the state where the conjugate pair of $E_3$ and $E_4$ are distinct. This is described in Figure 1. The whole situation speaks of the signature of an EP because the corresponding four eigenvectors coalesce there.

\begin{figure}[h!]
\centering
\begin{subfigure}{0.5\textwidth}
  \centering
  \includegraphics[width=0.95\linewidth]{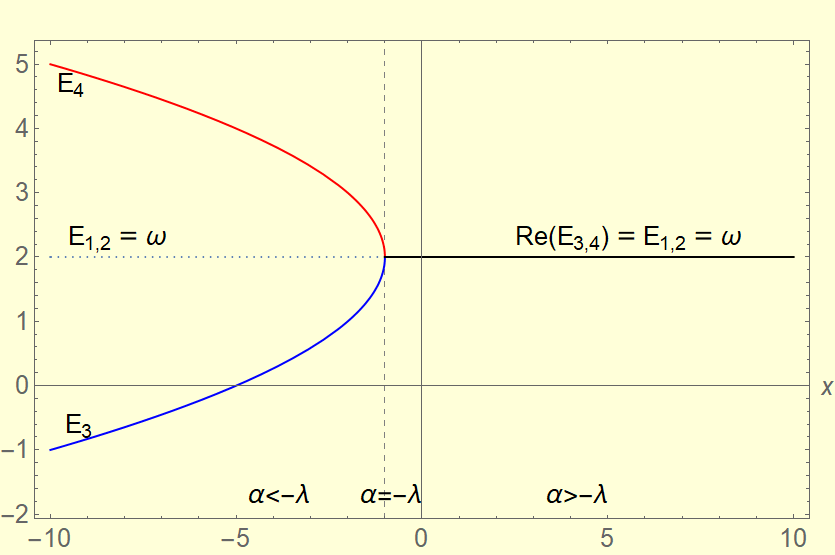}
  \caption{}
  \label{fig:sub1}
\end{subfigure}%
\begin{subfigure}{0.5\textwidth}
  \centering
  \includegraphics[width=0.95\linewidth]{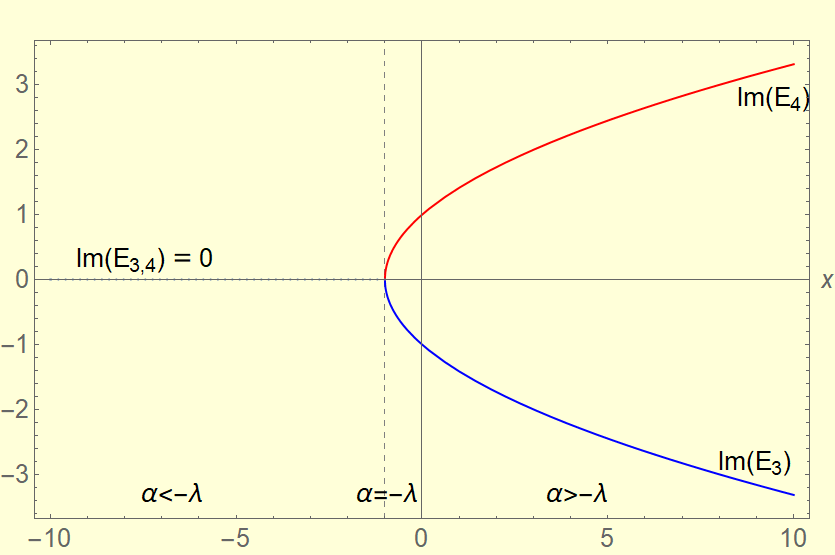}
  \caption{}
  \label{fig:sub2}
\end{subfigure}
\caption{(a) In the region $ \alpha <- \lambda $, $E_{1,2}$ are degenerate and real at the value $\omega$ as in (3.6), $E_{3,4}$ are also real but remain split for their respective values. The EP corresponds to the condition $ \alpha  = -  \lambda $. For $ \alpha  > - \lambda $, while $E_{1,2}$ continue to remain degenerate at $\omega$, $E_{3,4}$ become conjugate complex sharing their real part with $E_{1,2}$. (b) We plot Im$(E_{3,4})$ explicitly for $ \alpha  > -  \lambda $. For numerical estimates we took $\omega=2$ and $ \lambda  > \frac{1}{2}$.}
\label{fig:test}
\end{figure}

On the other hand, the situation is rather different for the complementary condition 

\begin{equation}
    \delta_{(+)} = \pm \sqrt{\gamma^2 - \alpha ^2 + 2 \alpha  \lambda  -  \lambda ^2}
\end{equation}
where $\delta_{(+)}$ is defined with respect to the positive-sign of $2 \alpha  \lambda  $ inside the square-root. Unlike those in (3.6), here $E_1$, $E_2$, $E_3$, $E_4$ have different values 
\begin{gather}
E_1= \omega -  \sqrt{2 \lambda ( \alpha  -  \lambda ) +2 \lambda \left| \alpha - \lambda  \right|},  \\  
E_2= \omega +  \sqrt{2 \lambda ( \alpha  -  \lambda ) +2 \lambda \left| \alpha - \lambda  \right| },\\
E_3= \omega -  \sqrt{2 \lambda ( \alpha  -  \lambda ) -2 \lambda \left| \alpha - \lambda  \right| },\\
E_4= \omega +  \sqrt{2 \lambda ( \alpha  -  \lambda ) -2 \lambda \left| \alpha - \lambda  \right| }
\end{gather}
Arguing similarly as in the previous case, we see that while for $ \alpha  >  \lambda $, all the energies are real

\begin{equation}
            \left ( \omega -  2\sqrt{ \alpha  \lambda  -  \lambda ^2}, \hspace{1mm}\omega +  2\sqrt{ \alpha  \lambda  -  \lambda ^2}, \hspace{1mm} \omega, \hspace{1mm} \omega \right)
\end{equation}
the transition to $ \alpha  <  \lambda $ results in $E_3$ and $E_4$ assuming complex values as follows from (3.10) and (3.11). Like before, for $ \alpha  =  \lambda $, all the four energies $E_{1,2,3,4}$ coalesce to the unique point $\omega$ while the corresponding four eigenvectors also coincide there. The presence of a bifurcation as evolving out of the EP is thus exhibited at the point $\omega$. See Figure 2.

\begin{figure}[h!]
\centering
\begin{subfigure}{0.5\textwidth}
  \centering
  \includegraphics[width=0.95\linewidth]{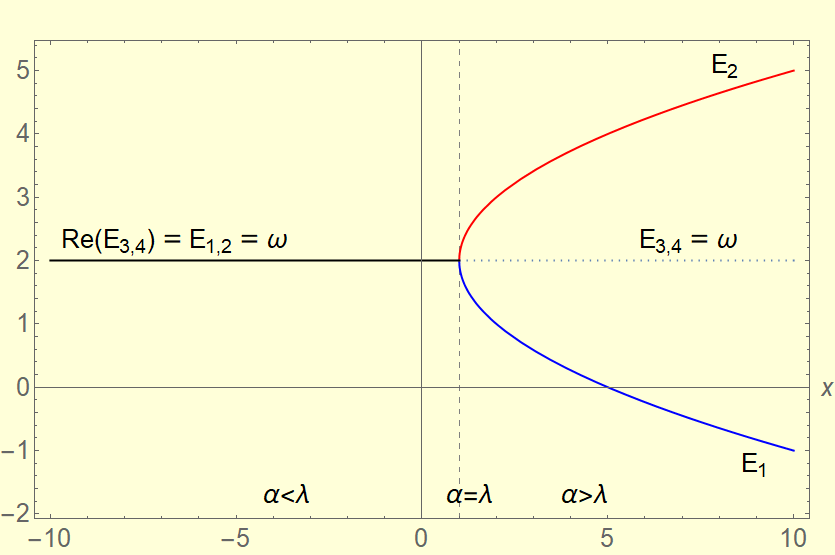}
  \caption{ }
  \label{fig:sub1}
\end{subfigure}%
\begin{subfigure}{0.5\textwidth}
  \centering
  \includegraphics[width=0.95\linewidth]{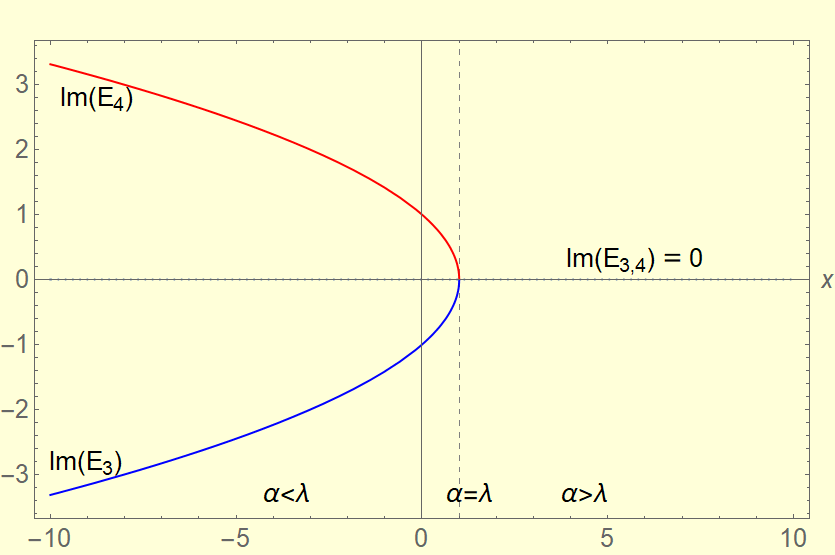}
  \caption{ }
  \label{fig:sub2}
\end{subfigure}
\caption{(a) In the region $ \alpha < \lambda $, $E_{3,4}$ are conjugate complex with Re$(E_{3,4})$ as well as $E_{1,2}$ having the common value $\omega$. The EP corresponds to the condition $ \alpha = \lambda $ where all the energies $E_{1,2,3,4}$ coalesce to $\omega$. Beyond it, i.e. when $ \alpha  >  \lambda $, $E_{1,2}$ continue to remain real but mutually unequal unlike the case in Figure 1 while $E_{3,4}$ stays fixed at $\omega$. (b) Im$(E_{3,4})$ is plotted. For numerical estimates we took $\omega=2$ and $ \lambda  = \frac{1}{2}$. }
\label{fig:test}
\end{figure}
Finally, it is to be stressed that with $\delta_{(+)}$ given by (3.5) and employing $ \alpha  = -  \lambda $, a fourth-order EP can be achieved which is formed by the coalescence of 4 eigenvectors. We do not expect profound physics to emerge out of it except that more complicated cases of degeneracy and types of EPs would be encountered.

We thus find that one can arrive at an EP while transiting either of the conditions $ \alpha  = \pm  \lambda $. Corresponding to them, we see that before an EP is reached, there is only a part-degeneracy of $E_1$ and $E_2$ as is clear from (3.6), and similarly for $E_3$ and $E_4$ as is evident from (3.12). The point to observe is that the behaviour of the system is sharply different for the two cases. In the first one, it is evident from (3.6) that as the phase transition occurs,  $E_3$ and $E_4$ gradually collapse to the EP $\omega$, $E_1$ and $E_2$ having already settled there. However, for the condition (3.12), a different picture emerges altogether in that both the pairs $(E_1, E_2)$ and $(E_3, E_4)$ are equidistantly split on the two sides of the point $\omega$, with different values of discontinuity resulting from the different signs of the inner square root in their respective expressions. At the transition point, the discontinuities wash out and all the four energies formally assume the coalescence value of $\omega$.

\section{Summary}

The study of nonhermitian quantum systems has attracted a good deal of attention over the past few years. The presence analysis deals with a nonhermitian Swanson-like model describing a coupled system of bosonic oscillators in which their mutual interactions are taken into account in a minimal way. We sketched the behaviour of the parametric interplay when these oscillators are assumed to be of the same frequency. Since a detailed treatment of the parameter interplay keeping all the elements present in the Hamiltonian will be out of bounds for practical reasons, we concentrated on specific constraints among the coupling. We are able to show that an EP as a nonhermitian degeneracy exists while the eigenstates undergo bifurcation. Indeed for the two sets of parametric conditions the existence of a bifurcation point is implied in the vicinity of the EP where the corresponding eigenvalues coalesce. The transition points have been pictorially depicted.

\section{Data availability statement}

All data supporting the findings of this study are included in the article.




\section{Acknowledgment}

Two of us (RG and SS) thank Shiv Nadar University for providing research fellowships. We are grateful to the anonymous reviewers for constructive criticisms and providing useful suggestions.

\end{document}